**Title:** Application of Deep Learning on Single-Cell RNA-sequencing Data Analysis: A Review
**Running title:** Deep learning in scRNA-seq data analysis

**Authors:** Matthew Brendel[1,2#], Chang Su[3#*], Zilong Bai[1], Hao Zhang[1], Olivier Elemento[2], Fei Wang[1*]
[1] Department of Population Health Sciences, Weill Cornell Medicine. Cornell University. New York, NY 10065, USA.
[2] Institute for Computational Biomedicine, Caryl and Israel Englander Institute for Precision Medicine, Department of Physiology and Biophysics, Weill Cornell Medicine. Cornell University, New York, NY 10065, USA.
[3] Department of Health Service Administration and Policy, Temple University, Philadelphia, PA 19122, USA.
# Matthew Brendel and Chang Su contributed equally to this work
*** Corresponding authors:**

**Chang Su,**
Department of Health Service Administration and Policy, Temple University, Philadelphia, PA 19122, USA.
Email: su.chang@temple.edu

**Fei Wang,**
Department of Population Health Sciences, Weill Cornell Medical College, New York, NY 10065, USA.
Email: few2001@med.cornell.edu

**Abstract**
Single-cell RNA-sequencing (scRNA-seq) has become a routinely used technique to quantify the gene expression profile of thousands of single cells simultaneously. Analysis of scRNA-seq data plays an important role in the study of cell states and phenotypes, and has helped elucidate biological processes, such as those occurring during development of complex organisms and improved our understanding of disease states, such as cancer, diabetes, and COVID, among others. Deep learning, a recent advance of artificial intelligence that has been used to address many problems involving large datasets, has also emerged as a promising tool for scRNA-seq data analysis, as it has a capacity to extract informative, compact features from noisy, heterogeneous, and high-dimensional scRNA-seq data to improve downstream analysis. The present review aims at surveying recently developed deep learning techniques in scRNA-seq data analysis, identifying key steps within the scRNA-seq data analysis pipeline that have been advanced by deep learning, and explaining the benefits of deep learning over more conventional analysis tools. Finally, we summarize the challenges in current deep learning approaches faced within scRNA-seq data and discuss potential directions for improvements in deep algorithms for scRNA-seq data analysis.

# Introduction

Since the first single-cell RNA sequencing (scRNA-seq) paper in 2009 and subsequent designation of "method of the year" a few years after, there has been a significant amount of effort to advance both the experimental and computational techniques used for the study of single-cell transcriptomes [1]. The benefit of scRNA-seq, compared to bulk RNA-seq, is the ability to interrogate thousands of individual cells simultaneously, thus revealing previously hidden heterogeneous cellular populations. scRNA-seq can then be used to answer biological questions related to developmental processes, understand complex and heterogeneous cellular or genetic changes based on treatment conditions or disease states, or identify novel cell types within a cellular population. Many popular packages, such as Seurat [2], Scanpy [3], Monocle [4] and Bioconductor OSCA (orchestrating single-cell analysis) [5], have been developed for a streamlined and reproducible analysis of scRNA-seq data. A pipeline for scRNA-seq analysis typically contains three steps (see Figure 1): (1) scRNA-seq data collection that produces a gene by cell matrix, of which elements are the raw gene expression read counts or unique molecular identifiers (UMIs), normalized to account for total genes captured for a particular cell either using standard approaches such as log or square root normalization, or more advanced approaches such as sctransform [6]; (2) data preprocessing including representation learning and dimensionality reduction, as well as optional doublet removal, cell-cycle variance removal, data imputation and denoising, and batch effect removal; and (3) downstream analyses, such as cell clustering, cell type annotation, and trajectory inference for discovery of cellular dynamic process along cells' development [7]. The result of this process can be used to answer biological questions of interest or determine unique features about the cellular populations that have been discovered.

Machine learning, a branch of artificial intelligence relying on mathematical and statistical principles, uses sets of data to build models that can perform specific tasks of interest and help accelerate or improve human decision making. In recent years, machine learning has successfully been used to analyze high-throughput omics data to improve upon the understanding of biological mechanisms of human health conditions [8, 9]. Conventional machine learning approaches usually require a significant amount of effort to develop a feature engineering strategy designed by domain experts, especially in the analysis of uncertain, heterogenous, and high-dimensional data like scRNA-seq data. As one of the latest and most popular advanced sub-categories of machine learning, deep learning provides a methodology that is more powerful in discovering latent, informative patterns from complex data and has achieved extraordinary improvements in computer vision and natural language processing (NLP) tasks. Importantly, compared to conventional machine learning, deep learning models can have thousands to millions of trainable parameters, which allow these models to uncover complex, non-linear patterns within the data in an end-to-end manner for improved analysis, specifically in the context of biology. In addition, deep learning models have a flexible architecture, which can be easily adjusted or assembled to adapt to solve different problems. Early evidence has demonstrated tremendous ability of deep learning in identifying underlying, informative patterns from scRNA-seq data, accounting for the heterogeneity present between scRNA-seq experiments, and noise and sparsity associated with scRNA-seq [10-12].

This review focuses on the use of deep learning in advancing key steps in the scRNA-seq data analysis. Extending on previous work [10-12], this review provides a comprehensive survey of deep learning in scRNA-seq data analysis. This review first provides an overview of deep learning, then introduces the most comprehensive list of deep learning models that have been used for various aspects of scRNA-seq data analysis, and finally, discusses limitations of these approaches and potential future directions in the field for improved scRNA-seq data analysis.

To narrow the scope of the paper, some aspects of scRNA-seq analysis have been excluded. Firstly, any discussion about sequencing read quality checks, read alignment, or quality checks for the alignment have been excluded, as deep learning is not involved in these procedures. Secondly, there is no discussion of RNA velocity-based downstream analyses, which involve identifying developmental transitions between cell types, including approaches such as DeepCycle [13] and VeloAE [14]. Since the input to the RNA velocity differs from that of standard scRNA-seq data analysis, which requires splicing information, this topic has been excluded. In addition, techniques such as Cobolt [15], scMM [16], and Schema [17], that combine information from multiple types of single-cell omics data have been excluded; this is to avoid providing extensive background on all different types of sequencing and antibody-based signal recognition approaches. Finally, studies that focused on simulating scRNA-seq data using deep learning, such as ESCO [18] and ACTIVA [19] are also excluded as they are not strictly necessary for scRNA-seq data analysis. More details of article inclusion and exclusion criteria can be found in the Supplementary Method and Supplementary Figure S1.

# Deep learning architecture in scRNA-seq data analysis

To differentiate machine learning from deep learning, we can refer to deep learning as the use of deep neural networks (DNNs) where "deep" describes the multilayer network structure. A deep feed-forward neural network (DFNN) is the most basic deep architecture by simply stacking layers of "neurons" (see Figure 2A). An artificial neuron is the basic computational unit of the DNNs, which takes the weighted summation of all inputs and feeds the result to a non-linear activation function, such as sigmoid, rectifier (i.e., rectified linear unit [ReLU]), and hyperbolic tangent (see Figure 2B), inspired by how human neurons work. A layer consists of a set of neurons and a DNN is built by stacking layers (see Figure 2A). In the basic design, a neuron receives information from all neurons of the previous layer with trainable weights while sending its output to the successor layer. Mimicking information flow in a human brain, the input information (i.e., gene expression profiles of the cells in scRNA-seq) flows from the input layer through the hidden layers and then the model generates an output at the last layer, i.e., the output layer. The large set of trainable weights of the neurons and the non-linear transformations enable the DNNs to capture underlying complex patterns of the data. Training of a DNN is the procedure of determination of these trainable weights that optimize model performance. In deep learning, the model training is typically done based on backpropagation, which mathematically transmits model prediction error in the reverse order of information flow from the output layer to the input layer to update model parameters or weights [20].

Based on the task of interest and the manner of model training, machine learning, and subsequently deep learning, can be grouped into three main categories: supervised learning, unsupervised learning, and semi-supervised learning. The standard DFNN (deep feed-forward neural network) is an architecture mainly used for supervised learning (see Figure 1A). In this scenario, the information available consists of a set of training data and the labels associated with each observation within the training set. The goal is to map the input data to a representation that can be used for tasks such as classification (for categorical labels) or regression (continuous labels). Semi-supervised learning, works when few data points have labels, using the limited labels to help inform the representation and label of the unlabeled data. Several scRNA-seq studies in this review use such technique, although it is not frequent.

There are several deep learning architectures suitable for unsupervised learning, which model data without any supervision and focuses on identifying underlying patterns from the data and are widely used in scRNA-seq data analysis, such as scRNA-seq data dimensionality reduction and cell clustering. The **deep autoencoder** (or autoencoder for simplicity) is a variant of the DFNN for unsupervised learning, which aims at learning compact representations of data while attempting to maximally preserve input data information (e.g., raw input gene expression in scRNA-seq) [21, 22]. A deep autoencoder typically consists of two components: an encoder and a decoder (see Figure 2C). The encoder is a DFNN that compacts data into a low-dimensional feature space at the so-called bottleneck layer. Then the decoder, with a mirror structure of the encoder, reconstructs the data in the original space from the low-dimensional representations derived by the encoder. Parameters of the autoencoder can be learned through minimizing such reconstruction errors using backpropagation. The learned low-dimensional representations of samples (i.e., cells in scRNA-seq data) are also called embeddings. Compared to those non-deep learning models like principal component analysis (PCA) that are components of well-established scRNA-seq data analysis software like Seurat [2], an autoencoder is capable of finding a non-linear manifold where the data lies [20].

To overcome pitfalls of autoencoders like overfitting, several modifications to the autoencoder structure have been proposed that contain specific benefits for scRNA-seq data (see Figures 2D-F). For instance, the denoising autoencoder (DAE) corrupts the input data slightly, by adding noise to a certain percentage of inputs, and then tries to rebuild the original input (see Figure 2D). In this way, model robustness in overcoming data noise is enhanced, and hence quality of the low-dimensional representation of samples (i.e., cells) learned from the scRNA-seq data [23, 24] is improved. This can be added on top of standard regularization strategies such as L1 and L2 regularizations of model weights.

Variational autoencoders (VAEs) are a type of generative model, as opposed to a discriminative model like the standard autoencoder. A VAE learns a latent representation distribution (such as Gaussian distribution), instead of a specific vector, which can be used to generate examples of cells' latent representations (see Figure 2E). Compared to the standard autoencoders, VAEs allow for reduced dimensionality, but also the quantification of uncertainty of the latent representation [25]. In addition, VAEs allow for a smoother latent representation of the data, which is beneficial when trying to understand relationships between cells at lower dimensions. For example, the smoothed low-dimensional representations can help improve accuracy in measuring distance between cells, when using metrics like Euclidean distance. The variational component of the optimization process acts as a regularization term for the autoencoder to improve generalizability to other data sources [26]. Typically, training of a VAE is based on the loss function composed of the reconstruction error (such as mean-squared error) and the Kullback–Leibler (KL) divergence between the latent distribution and an assumed prior distribution. In this context, VAEs can suffer from KL vanishing, or loss of informativeness for the latent representation (latent space exactly matches prior distribution). Modifications, such as the β-VAE and other variations on it [27], have been developed to address these issues and adapted for single-cell analysis. In addition, depending on the value of β, these models also have been shown to improve the disentanglement, or the independence of the latent dimensions, which can advance scRNA-seq data analysis. In addition, by involving an adversarial loss function, popularized by generative adversarial networks (GANs) [28] that have proven to be useful in synthetic data generation in other contexts, the VAEs can be described as an adversarial autoencoder [29].

Graph neural networks (GNNs) have successfully been applied to graph or network structured data analysis [30]. Typically, in each GNN layer, each node aggregates information from its local neighbors in the graph to update its representation (see Figure 2F). The graph autoencoder (GAE) is a novel modification of autoencoders by using GNN layers (see Figure 2F). In scRNA-seq data analysis, a cellular graph is usually built from the k-nearest neighbor (KNN) or shared nearest neighbor (SNN) strategies based on cells' gene expression profiles [2]. In this context, the GAE can be used to learn cell (i.e., node in the cellular graph) representations by incorporating cellular graph structure to decrease the noise of an individual cell. Figure 2F illustrates an example of GAE architecture for scRNA-seq analysis. Specifically, the encoder takes as input both gene expression read count matrix and cellular graph to generate cell representations, while the decoder(s) reconstructs the cellular graph structure (or both cellular graph structure and gene expression profile). There have also been more recent graph structures, using known protein-protein interaction (PPI) and cell-gene graphs, as prior knowledge, to improve scRNA-seq data analysis [63].

# Applications of deep learning in scRNA-seq data analysis

This section describes how deep learning is currently being used to improve key steps in scRNA-seq data analysis (see Table 1).

## scRNA-seq data imputation and denoising

An intrinsic pitfall of scRNA-seq is that as little as 6-30% of all transcripts are captured, based on the version of the chemistry used during sequencing and limited sequencing depth per cell [31]. Therefore, stochastically, cells will have what is known as "dropout" or the loss of all transcripts for a given gene [32], which is not biologically meaningful or accurate. From the data perspective, zero expression levels can be observed in the single-cell gene expression matrix, however some of them are "true" zeros, indicating the lack of expression of genes in specific cells, while unfortunately some others could be "false" zeros observed from genes that are expressed, i.e., dropout events, due to the low RNA capture rate (see Figure 1B). Therefore, when imputing missing values in scRNA-seq data, one must distinguish the "true" zeros and “false” zeros (see Figure 1B). This makes scRNA-seq data imputation more difficult than that of other biomedical data (such as clinical data), where missing values can be identified easily. Hence people also refer to the imputation procedure as scRNA-seq data denoising. It is important to note that denoising is not used in all deep learning-based approaches and therefore can be considered a potential component of the model, and benchmarking studies should be performed to see if it provides substantial benefits.

To account for the issue, conventional approaches [33-35] were proposed mainly focusing on imputing missing values based on correlated or similar genes or cells. However, they are usually computationally intensive and limited in capturing non-linearity in scRNA-seq data. To better address this issue, deep learning approaches have been developed for scRNA-seq data imputation and denoising [36-49]. Based on an idea similar to regression imputation [50], i.e., predicting missing values of target features (genes) using other features as predictors, DeepImpute (deep neural network imputation) [36] has been shown to be an effective approach for scRNA-seq data imputation using deep learning. Since DeepImpute only focuses on a subset of genes to impute (default 512), it can take advantage of the strength of the deep neural network but also reduces model parameters to make itself efficient and scalable. scIGAN (GAN for single-cell imputation) [37] leveraged a novel deep learning model, GAN (generative adversarial network). Specifically, scIGAN generates cells to impute dropout events, instead of using observed cells.

Other efforts that aimed at solving the scRNA-seq data imputation task use deep autoencoders. Intuitively, the reconstructed values by an autoencoder can be used to fill missing values in the original single-cell gene expression data. Based on such idea, a recent scRNA-seq analysis pipeline, scGMAI [38], has used an autoencoder for data imputation. Their experimental results on seventeen public scRNA-seq datasets demonstrated improvements of the autoencoder-based imputation in cell clustering task. SAVER-X [39] also used a standard autoencoder to denoise data. What makes SAVER-X unique is that the autoencoder was used to model the portion of expression of each gene that is predictable by other genes. Another innovation of SAVER-X is the incorporation of transfer learning framework. Particularly, the autoencoder can be pretrained using public cross-species (human and mouse) datasets, making it capable to transfer knowledge learned from mouse data to improve human data analysis.

In addition, some other studies combined the autoencoder architecture with parametric functions to facilitate imputation. DCA (deep count autoencoder) [40] used the zero-inflated negative binomial distribution (ZINB) noise model, which is effective at characterizing discrete, over-dispersed, and highly sparse count data, into the autoencoder architecture. Instead of directly reconstructing input data, DCA can produce three gene-specific parameters of ZINB, including mean, dispersion, and dropout probability, at the last layer of the autoencoder. After model training, the mean matrix from the output of the decoder can be used as a "denoised" version or imputed version of the original count matrix for downstream analysis. Yet, ZINB has its inherent shortcomings. As allowing three parameters for describing each data point, ZINB may be over-permissive to give a too high degree of freedom which may make the results unstable. To overcome this, ZINBAE (ZINB model-based autoencoder) [41] developed a ZINB autoencoder by introducing a differentiable function [51] to approximate the categorical data and a regularization term to control the ZINB. scSDAE (sparsity-penalized stacked denoising autoencoder) [42] leveraged a stacked DAE for scRNA-seq imputation with L1 loss to prevent overfitting. GraphSCI [43] combined the graph convolutional network (GCN), a type of GNN, with the standard autoencoder to model gene-gene co-expression relations and single-cell gene expression matrix, respectively. The incorporation of gene-gene co-expression relations as prior knowledge helps to alleviate bias in imputation and reduce impact of technical variations in sequencing.

It is worth noting that a notable benefit of deep learning in scRNA-seq data imputation and denoising is that there could be some non-linear relationships between certain genes. The deep architecture would allow for a more informed imputation strategy as compared to standard linear approaches. In addition, whether or not the zero-inflated negative binomial model is appropriate has been debated [52]. Finally, additional information, such as mapping relationships between genes in a graph structure, has been used for improved imputation.

## Doublet removal

The two main technologies used in single-cell isolation for downstream sequencing are microfluidic approaches, where cells are individually placed into oil droplets using microfluidic devices, and nanowell-based approaches, where tiny, patterned wells are created and individual cells are placed within each well [31, 53]. While these technologies have been improved and even commercialized over the past decade, errors can occur, in which more than one cell is captured within a droplet or well, i.e., so-called a "doublet". This can lead to improper interpretation of gene expression for a particular cell as the expression is a combination of multiple, and possibly different types of cells. This can happen if cells are not completely disassociated from one another after collection of the biological specimen.

To address this, single-cell doublet detection techniques have been developed. Typically, a doublet detection technique can be broken down into 3 main stages: doublet simulation, cell representation learning, and classifier training [54]. Solo [54] is a single-cell doublet detection model that leveraged the deep learning technique. For stage one, i.e., doublet simulation, Solo repeatedly took a random subset of cells (assumed to be singlets or single cells) and summed their UMIs, to generate $N$ different simulated doublets. For stage two, an unsupervised scRNA-seq data representation learning is engaged to embed these cells, singlets, and simulated doublets into a low-dimensional space. Specifically, Solo used the VAE-based representation learning model, scVI [55], to achieve the informative and robust cell representations. For stage three, Solo removed the decoder region and froze the weights for the encoder region. A set of fully connected layers were added to the end of the encoder, and then the model was trained to distinguish "singlet" and "doublet". Interestingly, scVI accounted for sequencing depth, which the authors state was a critical feature to include when running their model.

Traditional machine learning approaches for doublet detection, including Scrublet [56] and DoubletFinder [57], differ in the representation learning approach, usually PCA, as compared to a VAE, and in the way the authors identify doublets, relying on nearest neighbor approaches, compared to a neural network used in Solo. Interestingly, for Solo, the authors tested using both a VAE with KNN (K nearest neighbors) classifier and PCA with a neural network classifier, both of which performed worse in identifying doublets. This may highlight the need for both non-linear dimensionality reduction, to model the non-linear relationship between combinations of cells, and the need for a nonlinear classifier, as the latent space can still have non-linear relationships between singlets and doublets.

## Cell-cycle variance annotation

Gene expression can change as the cell moves along its normal cell cycle. The frequency by which cell types move between phases of the cell cycle varies due to many different factors [58], and can impact the expression of certain genes as a function of cycle. This change may add additional noise to downstream gene expression analysis and such uninformative variation between cells should be removed, or these changes may be useful information for downstream interpretation of sequencing data,. Typically, Seurat [2] used a cell scoring package, which can be used to regress out or subtract out the influence of cell cycle in the PCA latent space or explain variation among cells based on cell stage. Our literature search did find one study, Cyclum [58], which utilized the deep learning technique to account for cell cycle regression. Cyclum aimed at finding a nonlinear periodic function that encodes the cells' gene expression profiles to low-dimensional space and are sensitive to circular trajectories. To this end, it used a modified asymmetric autoencoder, which was composed of a standard encoder for representation learning and a decoder that uses a combination of cosine and sine as activation functions in the first layer and followed by a second layer for linear transformations. As a direct comparison to other linear methods (such as PCA), Cyclum showed superior performance in all datasets, measured using Hoechst staining of cells to identify cell cycle as ground truth labels. The test sets have a somewhat homogeneous cell population, so benchmarking on other datasets, with several different cell types, may be interesting to identify model performance, and improvement in subsequent downstream analyses.

## scRNA-seq data representation learning for dimensionality reduction

scRNA-seq data typically contains genome wide expression profiles of cells and hence has a very high feature space, making data analysis challenging due to the curse of dimensionality. The emerging term, scRNA-seq data representation learning, refers to the process of learning meaningful (information preserved) and compressed (low-dimensional) representations, or so-called embeddings, of cells based on their gene expression profiles and has been an essential intermediate step in single-cell analysis. It can not only advance other scRNA-seq data preprocessing procedures, such as doublet detection and cell-cycle variance annotation, but also benefit downstream analyses such as cell clustering, cell type annotation, and trajectory inference.

Early efforts in scRNA-seq data dimensionality reduction aimed at identifying a set of highly variable genes [2]. In addition, principal component analysis (PCA), which aims at determining principal components that can largely describe variance of the original data, has also been widely used to reduce dimensionality of scRNA-seq data. Though PCA is used in well-established software like Seurat [2], it cannot capture non-linear patterns in data and hence may harbor limitations when it comes to accurately reflecting the nature of cells. Due to their intrinsic ability to learn underlying, meaningful, and non-linear patterns from raw data [20], deep learning approaches [55, 59-88], especially the deep autoencoder and its extensions, have been effective techniques for scRNA-seq data representation learning and dimensionality reduction.

scScope [59] used an autoencoder to learn improved low dimensional representations of scRNA-seq data while simultaneously addressing dropout events. To this end, scScope introduced an imputer layer to generate a corrected input data based on the output of the decoder and re-sent it back to the encoder to re-learn an updated latent representation in an end-to-end manner. VAEs, which have shown the ability to disentangle latent representations or improve independence of latent dimensions [89], have demonstrated notable achievements in scRNA-seq representation learning. VASC (VAE for scRNA-seq data) [60] is an early effort that used VAE architecture with a a zero-inflated layer to account for dropout for scRNA-seq data dimensionality reduction. Compared to the traditional approaches, VASC resulted in better representations for very rare cell populations but also performed well on data with more cells and higher dropout rate. scVI (single-cell variational inference) [55] also used a VAE for scRNA-seq representation learning. It aggregated information across similar cells and genes to approximate latent distribution of the raw expression data but also accounted for batch effects. scDHA (single-cell decomposition using hierarchical) [62] leveraged an autoencoder combined with an ensemble of VAEs for learning informative representations of cells while preventing overfitting.

In VAEs, the modification of the prior distribution can be used to enhance the learned latent representation. scVAE (variational autoencoders for single-cell data) [64] utilized a Gaussian mixture model to model the latent representation instead of a standard normal. The Gaussian mixture enables the model to learn robust representations but also discover latent cluster structure simultaneously. scPhere [65] used the von Mises–Fisher (vMF) distribution to project data points onto the surface of a unit hypersphere and tested model variants that use hyperbolic space as the latent embedding [90]. In this way, scPhere decreased the crowding of points associated with normal VAE training and improved temporal information of data. In addition, there are modifications to loss function of VAEs to improve the disentangled representation. The basic VAEs, which typically use a KL loss (Kullback–Leibler divergence), may suffer from the issue of less informative representation, i.e., the learned representations are insufficient to represent the original data [91]. To overcome such issue, the DiffVAE [66] and MMD-VAE (maximum mean discrepancy VAE) [67] utilized a MMD loss instead of the traditional KL loss in the VAEs. DR-A (dimensionality reduction with adversarial variational autoencoder) [68] is a model that utilized a modified VAE, where the KL divergence component is replaced with two adversarial losses, one for latent representation and another for reconstruction. scRAE [69] builds upon this by modifying the adversarial autoencoder structure. Instead of sampling from a prior distribution and feeding that directly into the adversarial arm of the model, which is done in DR-A, the authors add a neural network after sampling from the prior distribution to be matched with the latent distribution generated from scRNA-seq autoencoder part of the model. The authors argue this form of regularization allows for a reduction in the bias associated with an assumed normal distribution, such as in DR-A, and shows this model outperforms several other approaches including DR-A. Kimmel [70] introduced a β-VAE to learn a disentangled representation of scRNA-seq data. While the author did see improvement in some downstream analyses such as identifying different cell conditions from the representation, others, such as cell type clustering had a decreased performance.

GAEs have also been used to model topology structure of relationships between cells in addition to the features (gene expression profiles) themselves, towards achieving better representations. Graph-DiffVAE [66] and scGAE (single-cell GAE) [71] are existing efforts in this context. Typically, they first constructed a cell-graph by connecting each cell to its K nearest neighbors based on gene expression profiles. Then it models and reconstructs the cell-graph and the gene expression matrix to learn low-dimensional representations of cells.

Model interpretability is a concern in deep learning. For scRNA-seq data, a common way for interpretable deep representation learning has been the use of prior domain knowledge, i.e., known relationships between molecules, like RNA and transcription factors, to modify standard neural networks. SCA (sparsely connected autoencoder) [72] used various forms of the autoencoder where the connections in the autoencoder are related to genes, transcription factors, miRNA targets, cancer-related immune signatures, and kinase specific protein targets. Additionally, other methods have leveraged similar known relationships, which allow for the construction of gene regulatory networks (GRNs). In the case of knowledge-primed neural networks (KPNNs) [86], the dimensionality reduction from the input, genes, to the output, phenotype, can be done by connecting nodes in one layer to the next that represent true relationships previously identified from large scale databases, such as SIGNOR (The SIGnaling Network Open Resource) [92] and TTRUST (Transcriptional Regulatory Relationships Unraveled by Sentence-based Text mining) [93]. GRNs can be reconstructed by analyzing the node weights across layers. Similarly, methods have utilized other forms of data representation using specific gene-gene correlations, to generate GRNs using more complex deep learning models, such as convolutional and recurrent neural networks [94]. In these settings, the supervised learning model can be thought of as a feature extraction method, that reduces the input feature space to a lower dimensional representation that can be used to predict whether there are specific interactions between genes.

More general pathway information is also useful to generate a more interpretable deep learning model. GOAE (gene ontology autoencoder) [73] used gene ontology (GO) [95] to determine the connections within an autoencoder. DeepAE [74] used an autoencoder and weights associated with each hidden unit to identify GO terms that are associated with high weighted genes. pmVAE (pathway module VAE) [75] encoded gene-pathway memberships for interpretable representation learning. Specifically, pmVAE contains a series of VAE subnetworks, each of which refers to a specific pathway module and only includes genes associated to this pathway. All pathway modules are combined to achieve global reconstruction of the input scRNA-seq data. VEGA (VAE enhanced by gene annotations) [76] performed a similar approach by masking genes such that genes within a certain gene module have similar contributions to a single latent dimension for the decoder. In addition, incorporating domain knowledge as a regularization term in the loss function to guide model training is another way to enhance interpretability. Rybakov et al. [77] injected GO into the loss function as a regularization term, such that genes associated with a certain pathway will be the only weights that contribute to the sum of a certain latent dimension. In LDVAE [78], the authors tried to improve interpretability of scVI by converting the decoder into a single linear layer, such that each gene can have a weight associated with each hidden unit in the latent space. While interpretability increases, there can be a decrease in performance, as now models are built based on known relationships and there could be some unknown relationships that are not modeled due to gaps in biological knowledge.

## Batch effect removal

Due to the stochastic nature of single cell sequencing, experiments done at different times, in different locations, using different reagents, using different technologies, or using different technicians, may have specific biases associated with that experiment that may influence sequencing results. To combat this, deep learning models [96-109] have been developed to learn a shared latent representation for these different experiments, that removes technical noise but keeps biological variation.

A common way to address this task is based on domain adaptation, which usually relies on GANs, an advanced branch of deep learning. Typically, a latent representation is generated using the autoencoder or its extensions, and then an adversarial training step is used in a discriminator module outside of the autoencoder to reduce difference in latent representations between batches. Following such an idea, iMAP [101] is a well-designed batch effect removal framework based on a autoencoder and GAN. Specifically, it used an encoder to produce batch-ignorant representation of cells and two generators to reconstruct the expression profile. Applied to tumor microenvironment datasets from two platforms, iMAP showed the capacity in taking advantage of powers of both platforms and identified novel cell-cell interactions using a non-deep learning approach, CellPhoneDB. In DAVAE (domain-adversarial and variational approximation) [97], a gradient reversal layer (GRL) was designed for domain adaptation to remove the batch effect. The scDGN (single-cell domain generalization network) framework [104] also used GRL. In contrast to other models, scDGN is trained in a supervised manner, aiming at maximizing the accuracy of cell type prediction while minimizing difference between batches. scGAN (single-cell generative adversarial network) [100] utilized a VAE architecture. The authors incorporated a discriminator module to predict batch from the data using an adversarial training. AD-AE (adversarial deconfounding autoencoder) [99] aimed to learn a confounder-free representation of data. The authors performed an adversarial optimization by adding an adversarial arm to the model to predict "confounders", such as batch and age. By alternating training by freezing the adversary arm weights and optimizing the loss by minimizing the reconstruction loss and maximizing the confounder loss and then freezing the autoencoder weights and minimizing the confounder prediction, the authors "remove" confounder information from the latent space. Pang and Tegnér [106] used BERT Transformer [110], an advanced attention-based neural network, as the encoder and an adversarial GAN based approach for batch alignment. SCALEX [98] incorporated a domain-specific batch normalization layer in the decoder of the VAE model to account for technical variations based on batches.

In addition to adversarial based approaches, there are also methods based on distribution matching, such as methods using different regularization terms like MMD (maximum mean discrepancy). BERMUDA (batch effect removal using deep autoencoders) aimed to match the latent representations learned by autoencoders between two batches [102]. Specifically, the autoencoder was performed on two batches separately. To overcome batch effects, the autoencoder was trained by optimizing a loss containing two components: a standard reconstruction loss and an MMD-based transfer loss between the latent representations of similar clusters from the two batches. trVAE (transfer VAE) [103] targeted at matching distributions across conditions. In the case of two conditions, the authors feed one condition into the encoder with the appropriate conditions associated with it. Then for the decoder the authors attach the opposite condition in the latent representation to transform the original condition feature matrix into the same space as the second condition. The MMD loss between the two conditions on the decoder region of the model was engaged to match distributions between different batches.

In addition, there are alternative ways to do batch correction. For example, the scScope pipeline [59] used a built-in batch correction layer in the DNN to performance batch correction. SMILE [96] utilized a contrastive learning framework [111], which forces each cell to be like itself plus a Gaussian noise while dissimilar to any other cells. scETM (single-cell embedded topic model) [105] used topic modeling to account for different batches and allow for some correction associated with batch-specific differences between cells. Specifically, it contains an encoder to infer cell type mixture and a linear decoder based on matrix tri-factorization.

## Cell clustering

One major goal of scRNA-seq analysis is to group the heterogeneous cell population into homogeneous sub-populations, such that cells within a sub-population are likely to have the same cell type or status. Clustering, an unsupervised learning approach, is a good fit to address this task. Typically, a clustering algorithm aims at identifying clusters, by minimizing dissimilarity within a given cluster while maximizing that between clusters. The well-established single-cell pipelines, such as Seurat [2] or Scanpy [3], use graph-based clustering methods such as Louvain [112] and Leiden [113] algorithms. Generally, they first build a cell-cell network using strategy like KNN (K nearest neighbors) based on gene expression profiles of cells, and then identified clusters by optimizing a measure such as "modularity" in Louvain [112], which measures cluster structure in the network (graph). In addition, the well-known K-means, which greedily adjusts clusters' centroids to optimize cluster structure, has also been widely used in scRNA-seq data analysis. Typically, the clustering algorithms take low-dimensional representations of cells as input, instead of raw gene expression profiles. In the deep learning setting [23-25, 114-123], the two steps, representation learning and clustering, can be done sequentially or simultaneously.

For the sequential modeling approaches, deep learning-based representation learning was performed first and followed by the classical clustering algorithms performed on the learned low-dimensional representations. scAIDE (single-cell autoencoder distance-preserved embedding network) [114] first provided a hybrid deep architecture for representation learning. Specifically, an autoencoder is used for imputation of the original input matrix, meanwhile a MDS (multidimensional scaling) encoder was used for dimensionality reduction. After that, scAIDE proposed a variant of K-means, called RPH-Kmeans, which utilized the LSH (locality sensitive hashing) technique [124] to tackle the data imbalance for clusters problem (i.e. different sized clusters) [114]. In addition, DUSC [23] made an extension to DAE for representation learning, i.e., DAWN (denoising autoencoder with neuronal approximator), which enables the model to automatically determine the number of latent features that are sufficient to represent the original gene expression data efficiently. The learned low dimensional representations were then used to identify clusters using an expectation-maximization (EM) algorithm [125]. scDMFK [115] also used DAE and combined with the fuzzy K-means algorithm to identify cell clusters. scCCESS [116] sampled the input data randomly to obtain multiple subsets. Then it learned low-dimensional representations in each subset using autoencoders and performed clustering subsequently. An ensemble clustering method was used to integrate clustering results in each subset to get the final one.

For the simultaneous modeling approaches, the models were designed in an end-to-end manner. Taking raw gene expression profiles as input, the data representation learning and clustering modules can be done automatically and these two modules can even improve each other in some advanced models. To achieve this, transfer learning is an intuitive option, which generally first pretrains a representation learning model, usually an autoencoder or its extensions, and then removes decoder and adds the pretrained encoder to another neural network for clustering. For instance, DESC [117] engaged a stacked autoencoder and pretrained it to learn low-dimensional representations of cells. After pretraining, the encoder was added to the neural network for cell clustering, in which batch effect can be removed over iterations in model training. CarDEC (count-adapted regularized deep embedded clustering) [118] is an advanced deep architecture that enables simultaneous batch effect correction, denoising, and clustering of scRNA-seq data. An innovation of CarDEC is that it treats the highly variable genes (HVGs) and lowly variably genes (LVGs) as different feature blocks. Specifically, it pretrained an autoencoder using HVGs, which were combined with LVG features for representation learning and clustering.

In addition, some authors designed hybrid deep architectures for joint representation learning and clustering. For instance, scziDesk (single-cell zero-inflated deep soft K-means) [119] learned data representation using ZINB (zero-inflated negative binomial) autoencoder while capturing non-linear dependencies between genes, and fed the learned representations to soft k-means clustering. The ZINB autoencoder and clustering module were trained jointly. GraphSCC [24] is a deep graph-based model for cell clustering. It contains three components: a DAE that encodes input gene expression profiles for preserving local structure, a GCN (graph convolutional network) encodes structural information of the cell-cell network, and a dual self-supervised module that connects the above two modules to learn informative latent representations of data and discover cluster structures. The low-dimensional representations learned by GraphSCC showed superior intra-cluster compactness and inter-cluster separability. scGNN (single-cell graph neural network) [120] is a hypothesis-free deep learning framework that integrates autoencoder, GNN, and left truncated mixed Gaussian modeling for scRNA-seq data analysis. scGNN performs imputation, representation learning, and clustering simultaneously, but also can produce a learned cell-cell interaction network.

All in all, both the sequential modeling approaches and simultaneous modeling approaches have shown improvement in cell clustering based on scRNA-seq data compared to the traditional non-deep clustering approaches. However, there has not been a direct comparison to show that performing the tasks sequentially or simultaneously has a strong impact on downstream analysis. This may be a future area of discussion and could be helpful when identifying which approach to use. In addition, tuning of the number of clusters based on the number of different cell types, and similarity between those cell types is something that is not fully investigated.

## Cell annotation

After cell clustering analysis, there is the need of interpreting or annotating the cell sub-populations, which is the so-called cell annotation. Traditionally, cell annotation can be done by identifying gene markers or gene signatures which are differentially expressed in the specific cell cluster and interpreting it manually [126]. However, such approaches are both labor- and resource-consuming. To address this, researchers are seeking deep learning approaches [127-147] that can handle this task with limited human supervision.

The supervised classification model, which can predict types or states of unlabeled cells based on labeled cells, is a good fit to address this task. For instance, scAnCluster [127] designed a hybrid deep model, which combined a cell type classifier with autoencoder for representation learning and clustering. JIND (joint integration and discrimination) [128] used a GAN style deep architecture, where an encoder is pretrained on classification tasks instead of using an autoencoder framework. The model is also able to account for batch effects. ItClust [129] engaged a transfer learning framework that pretrained model in source data to capture cell-type-specific gene expression information and then transferred model to identify and annotate clusters in the target data. scDeepSort [130] used an advanced GNN (graph neural network), GraphSAGE, to perform supervised classification for cell type annotation, accounting for cell interactions. AutoClass [131] used an autoencoder, where the output reconstruction loss is combined with a classification loss, for cell annotation with data imputation.

It is not uncommon to have only a subset of cells available for analysis with some level of annotation. In this context, semi-supervised learning, which can take full advantage of both labeled and unlabeled data to train a model, has been used in computational cell annotation. scANVI (single-cell annotation using variational inference) [132] is an extension of scVI [55] by incorporating semi-supervised learning to address cell type annotation with partial label information. scSemiCluster [133] learned cell labels using combination of unlabeled data and labeled data with an additional cluster compactness loss based on similarity matrix generation. scAdapt [134] used an adversarial training approach to perform semi-supervised cell type annotation. Specifically, it introduced the domain adaptation in DNN to include both adversary-based global distribution alignment and class-level alignment to preserve discriminations between cell clusters in the latent space. scAdapt has shown significance in cell annotation in simulated, cross-platforms, cross-species, and spatial transcriptomic datasets. scArches [135] used an architecture by concatenating nodes for new batches or datasets to existing autoencoder frameworks, to leverage information from other data sources. Moreover, in order for the utilization of the existing annotations to accelerate curation of newly sequenced cells, deep learning-based cell-querying approach has been proposed. Cell BLAST uses large scale reference databases with an autoencoder-based generative model to build low dimensional representations of cells, and use a developed cell-similarity metric, normalized projection distance, to map query cells to a specific cell type and allow for novel cell types to be identified [148].

Lastly, there is the situation where cell label information is very limited. To address this, there has been a study based on meta-learning to identify previously uncharacterized cell types. The meta-learning can train model to learn from models of known cell type classification to predict never-before-seen cell types. An existing effort in this context is the MARS [136], which used a DFNN (deep feed-forward neural network) as an embedding function to encode gene expression profiles. Under the meta-learning framework, the DFNN was shared by all experiments in the meta-dataset, which enables MARS to generalize to an unannotated experiment to address never-before-seen cell types.

## Trajectory inference

Biological questions can be answered by analyzing how cells change as they move from one cell type to another or one cell stage to another. Trajectory analysis in scRNA-seq is an approach to interrogate this type of question [7]. A "pseudotime" or developmental ranking of cells are established, such that the analysis seeks for how gene expression changes as a function of this time. The key process that is used for many approaches is transforming a latent representation of the model into a graph structure. Next, the model usually requires a start cell, which in developmental analyses is usually one with some "stem-like" marker. The algorithms developed traverse the graph, usually the novel component of most algorithms, to find a path from the start cell to several terminal states. Standard scRNA-seq data analysis tools that provide trajectory inference include Scanpy [3], Monocle [4], VIA [149], and Palantir [150], etc. To date, these tools have been using traditional methods like PCA for data dimensionality reduction for inferring trajectories. While approaches like VIA claimed that dimensionality reduction is not a necessary step for their algorithm, there remains the comparison between linear and nonlinear approaches for dimensionality reduction in this task. VITAE (variational inference for trajectory by autoencoder) [151] is an existing effort that uses deep learning to advance trajectory inference. Specifically, VITAE combined a VAE for latent representation learning with a hierarchical mixture model to represent the trajectory. The use of a deep learning model, VAE, enables VITAE to recognize non-linear patterns in data and adjust for confounding covariates to integrate multiple datasets at scale.

## Discussion: open issues and future directions

In this review, we have investigated how deep learning has been incorporated to advance different elements of scRNA-seq data analysis. Despite the promising results obtained using the deep learning techniques, there remain challenges in the field that need to be solved.

### Need of benchmarking studies

One of the most pressing needs, especially for the deep learning approaches developed for scRNA-seq analysis, are benchmarking studies. Most of the papers published using deep learning approaches compared performance to other standard methods but didn't go into great depth when comparing across different types of deep learning models. Single-cell experiments can be vastly different, with tissue samples that contain known cell types, such as in the pancreas (alpha cells, beta cells, delta cells, etc.) or from much more complex tissues, such as in diseases such as cancer or COVID, where there are many different cell types, and variations of cell types present within the tissue sample. However, most methods claimed superior performance only based on a set of example datasets from specific single-cell experiments. What's more, it is difficult to assess, with the vast number of approaches that have been developed, whether a certain regularization term or added preprocessing step is essential for a particular scRNA-seq data analysis. Therefore, to overcome the above issues, one potential way would be to better understand when these deep learning models fail or what the limitations are for these approaches. Understanding the types of deep learning approaches and model structures that can be beneficial in some cases as compared to others would be very important for (1) developing new approaches to handle these shortcomings and (2) guiding the field as to what methods perform better under specific conditions. In addition, another major improvement in the field would be the human cell atlas, i.e., the aggregation of many different human single cell expression data across many institutions to cover all major organ systems within the body. This will allow for large amounts of annotated scRNA-seq data, from multiple institutions. This collection of data can allow for more comprehensive benchmarking studies, as a dataset for standardized model evaluation, similar to that of ImageNet or CIFAR10 for computer vision algorithm developers. Fortunately, recent work is moving in this direction, as a group has just tested several batch correction approaches using an atlas level amount of single cell data and another group has tested 45 different single cell trajectory inference approaches on 110 different single cell datasets and proposed guidelines for method selection [7, 152].

## Integrative analysis of multiple datasets

While deep learning has been involved in continuously increased scRNA-seq data analysis studies, they usually suffer from limited available information of single dataset, on the order of several tens of thousands of single cells, for the analyses. At this point, it may be difficult to identify substantial amounts of rare cell populations and characterize how these rare cell populations change under varying disease states. These datasets are orders of magnitude smaller than datasets in computer vision tasks where deep learning has achieved significant improvements. For example, most deep learning models in computer vision are pretrained on ImageNet, which contains 1.2 million images split between 1000 different classes. With the increasing availability of scRNA-seq data, the use of these smaller datasets for computational analyses may be changing. Recent work by Sikkema et al, uses a combination of 46 different datasets with 2.2 million cells to analyze lung tissue across healthy and diseased patients [153]. The authors specifically did a benchmarking step to identify the appropriate single cell integration approach to use for their dataset, and found that the deep learning method, scANVI, outperformed all other methods, including the standard pipeline approach of Seurat. In addition, this was further validated in a large-scale benchmarking dataset [152], showing that two out of the three top performing methods were deep learning approaches. The authors suggest that standard approaches for data integration, such as Harmony, work best when biological complexity is small but are outperformed by deep learning approaches in more complex settings [152]. Additionally, the deep learning use of transfer learning, similar to approaches such as scArches, can be used to save the information gained from large-scale training sets, to additional researchers that do not have access to such large and diverse datasets. This idea of large-scale model training and transfer learning to fine-tune the model are key aspects of deep learning and a potential future direction in the field of scRNA-seq computational analysis. The field of scRNA-seq is continuing to embrace the concept of open-source data sharing, and new toolkits, such as scverse (https://scverse.org), look to provide a unified framework for doing these large-scale scRNA-seq analyses. Information gathered in these analyses, on top of other large scale data collection efforts such as TCGA, can be utilized to better understand how cellular changes correlate with disease [154]. In addition, for datasets where patient scRNA-seq and additional disease related information, such as the HPAP-DB dataset [155], information beyond transcriptome data can be used to identify how distinct cellular changes affect clinical phenotypes [156].

## Knowledge-enhanced deep modeling

As the field of deep learning has advanced, the "deep" architectures being developed have become more complex and more "black-box" like, in other words it is difficult to understand and interpret how the models work. To make deep learning useful for clinicians, and biology in general, interpreting deep learning models has been an active area of research. In addition, the "deep" architectures may result in the overfitting issue if the models developed are too complex and hence focus on limited details of the data. Meanwhile, the heterogeneous cell populations and the high dimensionality of gene expression profiles challenge the modeling training, potentially leading to underfitting, such that the models developed are not capable of sufficiently capturing patterns within the data. In this context, incorporating biomedical domain knowledge has been a desirable option to account for those issues in data analysis. To date, there have been several existing studies [72, 75, 77, 78] that developed knowledge-enhanced deep learning models for scRNA-seq analysis. Though these models have gained notable improvement in specific application areas, there remains considerable room for improvement as the knowledge used is limited to specific resources like the GO (gene ontology) knowledgebase. In addition, today's biomedical knowledge graphs (BKGs) [157-159] have been an important biomedical resource that store comprehensive knowledge in biology and medicine and have been engaged to improve omics data analysis[159-162]. Generally, a BKG is a type of biomedical knowledge base with a graph/network structure where nodes are a set of biomedical entities (e.g., diseases, drugs, genes, biological processes, etc.) and edges between nodes/entities are relations linking the biomedical entities (e.g., drug-treats-disease, disease-associates-gene, drug-interacts-drug, etc.) [157, 158, 163]. The BKGs have been used to interpret findings from omics data analysis through BKG query. For instance, Santos et al. [159] developed a clinical knowledge graph (CKG) platform, which enables clinically meaningful queries for automated proteomics data analysis, knowledge mining, and visualization. Doddahonnaiah et al. [161] used a BKG derived from literature to augment the annotation and interpretation of scRNA-seq data. The gene-cell type associations in their BKG were used to categorize cell clusters identified by scRNA-seq data. In addition, researchers have been seeking to develop BKG-guided machine learning and deep learning models to improve scRNA-seq data analysis. In their recent work, Cao and Gao [162] developed a deep learning model for multi-omics single-cell data integration and regulatory inference. Specifically, a graph VAE was used to learn feature embeddings from a prior knowledge-based guidance graph (a specific BKG), which were then fed to the omics VAE to reconstruct omics data via inner product with cell embeddings. In this way, unpaired multi-omics single-cell data such as scRNA-seq, scATAC-seq, and snmC-seq (with non-overlapped samples and features) can be projected to the shared cell embedding space.

## Integrative modeling with multi-omics data

The ever-improving single-cell isolation and barcoding techniques have been producing diverse omics data at single-cell level, such as genetics, genomics and transcriptomics, and proteomics [164]. On the other hand, integrative analyses of multi-omics data at the bulk level [165-167] have shown the promise to provide a comprehensive understanding of molecular mechanisms to accelerate biology and medicine, as it provides the route to study molecular processes from multiple angles. Compared to traditional machine learning methods, deep learning has demonstrated its superiority in bulk multi-omics data analysis [168-170], due to the capacity in capturing informative latent features from the high-dimensional heterogeneous multi-omics feature space, and the flexible architecture that can model each modality separately using small DNNs (e.g., autoencoders) and combine them later to aggregate information extracted from each modality appropriately to learn a joint representation [171]. Drawing on the success in bulk multi-omics data, integrating scRNA-seq data with other single-cell omics data as well as multi-omics data at bulk level using deep learning may help provide a better and deeper understanding of the biological mechanisms. While there have been many successes in multi-omics data integration [172-176], there remains specific and distinct challenges, for both joint-modality single cell sequencing, such as CITE-seq [177], and the integration of single modality single cell omics sequencing data. For joint modality sequencing, to leverage both datasets simultaneously, most methods employ a method of joint representation learning, or finding a shared latent representation of the data for all modalities. One challenge with this type of joint sequencing is that there can be an increase in noise and sparsity in the data, compared to scRNA-seq data using one modality [178]. In addition, it is difficult the balance of both modalities during the embedding process, and it is possible that some modalities can dominate the downstream embedding tasks leading to the reduction of biological variability that exists within one modality. Finally, there are inherent biases [179] between different institutions, making joint learning more challenging when generalizing across institutions. Additionally, joint sequencing models are much less frequent than single modality sequencing methods, so an important direction for analysis is to develop methods to integrate two different modalities with unique cell populations. In this setting some goals would be to predict the expression of one modality from another or identify cells in the same cellular state across different modalities. This remains a significant computational challenge, as highlighted by the 2021 NeurIPS single cell challenge. Several methods were developed in this challenge as well as outside, but more work can be done to improve overall performance and more work can be done to improve multi-omic analysis when unique or rare cell populations are in one technology, but not present within another.

Spatially resolved transcriptomics (SRT) is a new approach to single cell analysis that preserves the spatial relationship of RNA-sequencing within a tissue. While SRT has the advantage of spatial resolution, the major technology currently on the market, the 10x Visium platform, currently generates 50 micron spots that are pooled for analysis, losing the ability to identify the transcriptome of a single cell. There are other approaches that aim to improve the resolution, such as Slide-Seqv2 [180], but these too have drawbacks such as limited ability to detect low expressing genes compared with scRNA-seq methods [181]. It is therefore important to realize that scRNA-seq can act complimentary to the SRT technology. Firstly, SRT will require unique computational and deep learning algorithms, separate from scRNA-seq. For example, a method PASTE [182], shows that scRNA-seq methods are insufficient to properly analyze SRT data. In addition, cell-cell communication networks can be elucidated using newly developed algorithms [183]. However, scRNA-seq currently can provide unique gene information that has been leveraged during SRT analysis. For example, DestVI uses a reference scRNA-seq dataset to deconvolve or attempt to identify unique cell types within a given SRT spot [184]. In addition, work has been done to jointly embed seqFISH data and an scRNA-seq atlas, to annotate specific cell types in the seqFISH dataset [185]. Therefore, with current SRT spatial resolution constraints and detection limitations, SRT and scRNA-seq can act synergistically. Additionally, the autoencoder structures used in the context of scRNA-seq and in this review can also be components used within SRT analysis.

## Golden standard pipeline

We have discussed deep learning applications in steps in scRNA-seq data preprocessing, including data imputation, representation learning, doublet removal, batch effect removal, and cell cycle regression, and scRNA-seq data downstream analyses, such as cell clustering, cell annotation, and trajectory inference. However, there are several steps in the pipeline that we have discussed, such as doublet detection and imputation, not always used for analysis. The well-established software like Seurat and Scanpy do allow users to customize the analysis pipeline according to the application scenarios. Efforts like scAEspy [186] and sfaira [187] also built deep learning-based scRNA-seq data analysis pipelines. It will be important to perform thorough comparisons to validate (1) the need for each of these steps, (2) the better way to arrange them in the analysis pipeline, and (3) how deep learning impacts these steps to advance the whole analysis pipeline. There should be systematic effort to determine critical steps in the scRNA-seq analysis pipeline to assure that methods are being developed for critical steps in the analysis.

## Conclusions

scRNA-seq has been a critical technique to study cell level gene expression. Deep learning, a powerful artificial intelligence technique that has shown significant capacity in big data mining and outperforms the conventional machine learning, has now firmly been introduced in scRNA-seq data analysis. Specifically, deep learning has been involved in key steps to advance scRNA-seq data analysis. Notable achievements have been gained through the use of deep learning techniques compared to the traditional data analysis methods. By carefully reviewing and comparing existing applications of deep learning in scRNA-seq data analysis, we summarize the challenges the current deep learning applications are faced with and discuss potential future directions in this field.

## Competing interests

The authors have declared no competing interests.

## CRediT Author Statement

Matthew Brendel: Conceptualization, Writing - Original Draft, Writing - Review and Editing, Fei Wang: Conceptualization and Supervision, Chang Su - Writing - Original Draft, Writing - Review and Editing. Zilong Bai, Olivier Elemento and Hao Zhang: Writing - Review and Editing.

## Acknowledgments

FW would like to acknowledge the support from NSF 1750326, NIH R01MH124740 and RF1AG072449.

**Table 1.** A summary of the selected studies in this review

| Author Name, Publication Year | Category | Model Name | Model Type | Code Availability | Technical Advancement |
|---|---|---|---|---|---|
| Arisdakessian et al., 2019 [36] | Imputation and Denoising | DeepImpute | AE | https://github.com/lanagarmire/deepimpute (Python) | Utilizing correlated genes to impute missing values using autoencoder |
| Xu et al., 2020 [37] | Imputation and Denoising | scIGAN | GAN | https://github.com/xuyungang/scIGANs | Use KNN of a set of boundary equilibirum GAN-generated cells for a certain cell type to perform imputation |
| Yu et al., 2021 [38] | Imputation and Denoising | scGMAI | AE | https://github.com/QUST-AIBBDRC/scGMAI | Output of autoencoder with softplus activation functions used as imputed representation for further dimensionality reduction with FastICA and clustering with GMM |
| Wang et al., 2019 [39] | Imputation and Denoising | SAVER-X | AE | https://github.com/jingshuw/SAVERX | Novel empirical bayesian shrinkage approach to predicting imputed values from autoencoder output based on gene-gene relationships |
| Eraslan et al., 2019 [40] | Imputation and Denoising | DCA | AE | https://github.com/theislab/dca | Zero-inflated negative binomial loss for denoising |
| Tian et al., 2021 [41] | Imputation and Denoising | ZINBAE | AE | https://github.com/ttgump/ZINBAE | Gumbel softmax applied to dropout matrix of decoder output and zero-inflated negative binomial loss for denoised data representation |
| Chi et al., 2020 [42] | Dimensionality Reduction | scSDAE | DAE | https://github.com/klovbe/scSDAE | Stacked denoising autoencoder with L1 penalty only for values with 0 to induce sparsity into output |
| Rao et al., 2021 [43] | Imputation and Denoising | GraphSCI | AE \| GAE | https://github.com/biomed-AI/GraphSCI | Use gene-gene network derived from a thresholded pearson correlation calculation for improved imputation |
| Huang et al., 2020 (Preprint) [44] | Imputation and Denoising | SAVERCAT | VAE | - | Use highly variable genes to train conditional VAE, then use the learned parameters to denoise retrain the decoder using the entire set of genes for downstream analysis |
| Li et al., 2021 [45] | Imputation and Denoising | SEDIM | AE \| DFNN | https://github.com/li-shaochuan/SEDIM | Use learning algorithm to find optimal hyperparameters for model generation to perform imputation |
| Xu et al., 2021 [46] | Imputation and Denoising | AdImpute | AE | - | Use MSE on autoencoder output and imputed values from DrImpute in additon to standard autoencoder training |
| Xu et al., 2021 [47] | Imputation and Denoising | GNNImpute | GAE | https://github.com/Lav-i/GNNImpute | Use graph attention autoencoder to perform imputation |
| Gunady et al., 2019 (Preprint) [48] | Imputation and Denoising | scGAIN | GAN | https://github.com/mgunady/scGAIN | Concatenate mask of droput values and original count matirx with randomly intialized values, use hint generator to perturb original mask, and use adversarial training to predict which values in imputed cell representation are real or fake |
| Badsha et al., 2020 [49] | Imputation and Denoising | LATE/TRANSLATE | AE | https://github.com/audreyqyfu/LATE | Autoencoder with MSE for non-zero input values and transfer or learned weights to other datasets |
| Bernstein et al. 2020 [54] | Doublet Removal | Solo | VAE | https://github.com/calico/solo | scVI model used for dimensionality reduction with for doublet vs singlet embedding and neural network for classification of doublets |
| Liang et al. 2020 [58] | Cell-cycle variance removal | Cyclum | AE | https://github.com/KChen-lab/Cyclum | Circular activation functions in decoder to identify circular latent structures and subsequently cell cycle structure |
| Deng et al., 2019 [59] | Dimensionality Reduction | scScope | AE | https://github.com/AltschulerWu-Lab/scScope | Introduce the autoencoder output recurrently to impute missing values and improve latent representation |
| Wang et al., 2018 [60] | Dimensionality Reduction | VASC | VAE | https://github.com/wang-research/VASC | Modeling the data as zero-inflated (gumbel distribution) in decoder using VAE |
| Cho et al., 2018 [61] | Dimensionality Reduction | net-SNE | DFNN | https://github.com/hhcho/netsne | Apply t-SNE loss function to neural network |
| Lopez et al., 2018 [55] | Dimensionality Reduction | scVI | VAE | https://github.com/YosefLab/scvi-tools | Cell specific scaling of counts based on size factor for cell that is modeled into VAE |
| Tran et al., 2021 [62] | Dimensionality Reduction | scDHA | AE \| VAE | https://github.com/duct317/scDHA | Non-negative weights for non-negative kernel autoencoder for feature selection and multiple decoders in VAE for the stacked bayesian autoencoder for feature representation |
| Li et al., 2021 [63] | Dimensionality Reduction | scGSLC | GCN | https://github.com/sharpwei/GCN_sc_cluster | Use protein-protein interaction network to perform dimensionality reduction for improved clustering |
| Grønbech et al., 2020 [64] | Dimensionality Reduction | scVAE | VAE | https://github.com/scvae/scvae | Introduction of a gaussian mixture prior for the VAE training |
| Ding et al., 2021 [65] | Dimensionality Reduction | scPhere | VAE | https://github.com/klarman-cell-observatory/scPhere | Spherical or hyperbolic embedding to improve clustering and latent representation of single cells |
| Bica et al., 2020 [66] | Dimensionality Reduction | DiffVAE/GraphVAE | VAE | https://github.com/ioanabica/DiffVAE | VAE and GraphVAE framework for scRNA-seq analysis with InfoVAE model |
| Zhang 2019 (Preprint) [67] | Dimensionality Reduction | MMD-VAE | VAE | https://mmd-vae.hi-it.org/ | Replace KL divergence term with MMD for VAE training |
| Lin et al., 2020 [68] | Dimensionality Reduction | DR-A | AAE | https://github.com/eugenelin1/DRA | Adversairal loss on reconstructed output and latent space of the variational autoencoder |
| Mondal et al., 2021 [69] | Dimensionality Reduction | scRAE | AAE | https://github.com/arnabkmondal/scRAE | Uses a P-GEN network, or neural network to reduce the bias of the regularization term for the autoencdoer latent representation in VAE or AAE framework |
| Kimmel, 2020 [70] | Dimensionality Reduction | | VAE/β-VAE | - | Use β-VAE for disentangled representation of single cells generating more interpretable latent representations |
| Luo et al. 2021 [71] | Dimensionality Reduction | scGAE | GAE | https://github.com/ZixiangLuo1161/scGAE | Using graph attention encoder for dimensionality reduction |
| Alessandri et al., 2021 [72] | Dimensionality Reduction | SCA | AE | https://github.com/kendomaniac/SCAtutorial | Using known relationships of genes with transcription factors, kinases, and miRNA to model network connections for autoencoder |
| Peng et al., 2019 [73] | Dimensionality Reduction | GOAE | AE | - | Use prior knowledge gene ontology terms to impact the connection between layers for the autoencoder |
| Zhang et al., 2020 [74] | Dimensionality Reduction | DeepAE | AE | https://github.com/sourcescodes/DeepAE | Use weights from neural network to generate gene ontology terms for hidden representation dimensions |
| Gut et al., 2021 (Preprint) [75] | Dimensionality Reduction | pmVAE | VAE | https://github.com/ratschlab/pmvae | Ensemble of VAEs each with a pathway specific set of genes for more interpretable single cell representation |
| Seninge et al., 2021 [76] | Dimensionality Reduction | VEGA | VAE | https://github.com/LucasESBS/vega-reproducibility | Use mask on linear decoder weights to improve interpretation based on gene database |
| Rybakov et al., 2020 (Preprint) [77] | Dimensionality Reduction | | AE | https://github.com/theislab/intercode | Use pathway databases, such as MSigDB, to induce regularization into model for improved interpretability |
| Svensson et al., 2020 [78] | Dimensionality Reduction | LDVAE | VAE | https://github.com/YosefLab/scvi-tools | Restrict decoder of scVI to linear layer for improved interpretability |
| Zhou et al., 2021 [79] | Dimensionality Reduction | SCDRHA | GAE | https://github.com/WHY-17/SCDRHA | Uses output of DCA as input for Graph attention autoencoder |
| Wang et al., 2021 [80] | Dimensionality Reduction | scCDG | DAE \| GAE | https://github.com/WHY-17/scCDG | Graph Autoencoder on latent representation from denoising autoencoder |
| Buterez et al., 2021 (Preprint) [81] | Dimensionality Reduction | CellVGAE | GAE | https://github.com/davidbuterez/CellVGAE | Use variational graph attention autoencoder for dimensionality reduction |
| Ciortan et al., 2021 [82] | Dimensionality Reduction | graph-sc | GAE | https://github.com/ciortanmadalina/graph-sc | Input cell-gene graph into graph autoencoder for dimensionality reduction |
| Ciortan et al., 2021 [83] | Dimensionality Reduction | contrastive-sc | DFNN | https://github.com/ciortanmadalina/contrastive-sc | SimCLR loss using two different dropout representations of the same cell for self-supervised contrastive learning |
| Lukkassen et al., 2020 [84] | Dimensionality Reduction | resVAE | VAE | https://github.com/lab-conrad/resVAE | Mask out latent representation based on known cell type or other meta data |
| Prince et al., 2019 (Preprint) [85] | Dimensionality Reduction | HD Spot | AE | - | Genetic algorithm to optimize autoencoder hyperparameters and converting the encoder to a classifier to perform SHAP for improved interpretability of gene importance for different classes |
| Fortelny et al., 2020 [86] | Dimensionality Reduction | KPNN | DFNN | https://github.com/epigen/KPNN | Control node connections in neural network based on known biological pathways |
| Gold et al., 2019 [87] | Dimensionality Reduction | SSCA/SSCVA | AE \| VAE | - | Use known gene sets to control node connections in autoencoder |
| Yu et al., 2021 [88] | Dimensionality Reduction | MichiGAN | VAE \| GAN | https://github.com/welch-lab/MichiGAN | Use β-TCVAE for disentangled representation of single cells generating more interpretable latent representations |
| Xu et al., 2021 [96] | Batch Effect Removal | SMILE | DFNN | https://github.com/rpmccordlab/SMILE | Contrastive learning loss (NCE) for the integration of multiple datasets |
| Hu et al., 2021 [97] | Batch Effect Removal | DAVAE | VAE | https://github.com/jhu99/davae_paper | Gradient reversal layer for adversarial training to perform data integration |
| Xiong et al., 2021 (Preprint) [98] | Batch Effect Removal | SCALEX | VAE | https://github.com/jsxlei/SCALEX | Use decoder-based domain-specific batch normalization for multi-source data integration |
| Dincer et al., 2020 [99] | Batch Effect Removal | AD-AE | AE | https://gitlab.cs.washington.edu/abdincer/ad-ae | Adversarial training of autoencoder for multiple different confounders including age and batch to learn deconfounded latent representation |
| Bahrami et al., 2021 [100] | Batch Effect Removal | scGAN | VAE | https://github.com/li-lab-mcgill/singlecell-deepfeature | Adversarial training of VAE with categorical (batch) or continuous (age) variables for data integration |
| Wang et al., 2021 [101] | Batch Effect Removal | iMAP | AE \| GAN | https://github.com/Svvord/iMAP | Two step integration using (1) content loss and (2) random walk MNN-based GAN model |
| Wang et al., 2019 [102] | Batch Effect Removal | BERMUDA | AE | https://github.com/txWang/BERMUDA | Integration of cluster pairs between batches identified using MetaNeighbor with MMD regularization |
| Lotfollahi et al., 2020 [103] | Batch Effect Removal | trVAE | VAE | https://github.com/theislab/trVAE | Conditional VAE with MMD regualrization in latent space |
| Ge et al., 2021 [104] | Batch Effect Removal | scDGN | DFNN | https://github.com/SongweiGe/scDGN | Semi-supervised learning with domain adaptation using gradient reversal layer |
| Zhao et al., 2021 [105] | Batch Effect Removal | scETM | VAE | https://github.com/hui2000ji/scETM | Interpretable decoder using matrix tri-factorization (Topic Modeling) |
| Pang and Tegnér, 2020 (Preprint) [106] | Batch Effect Removal | | BERT Transformer | - | Use transformers for encoder and decoder |
| Zou et al., 2021 [107] | Batch Effect Removal | DeepMNN | DFNN | https://github.com/zoubin-ai/deepMNN | Use Residual Network to perform batch correction on predetermined MNN pairs of cells using highly variable genes |
| Wang et al., 2021 [108] | Batch Effect Removal | HDMC | AE | https://github.com/zhanglabNKU/HDMC | Contrastive loss using MetaNeighbor-identified similar clusters between batches for improved batch correction |
| Yu et al., 2021 [109] | Batch Effect Removal | CBA | AE | https://github.com/GEOBIOywb/CBA | Integrate pre-defined matching cell clusters from two domains using a two-stream autoencoder network, which uses concatenation of latent representation within and between streams |
| Xie et al., 2020 [114] | Cell Clustering | scAIDE | AE \| DFNN | https://github.com/tinglabs/scAIDE | MDS encoder for improved autoencoder dimensionality reduction and rph kmeans for improved clustering of different sized clusters |
| Chen et al., 2020 [115] | Cell Clustering | scDMFK | AE | https://github.com/xuebaliang/scDMFK | Simultaneous dimensionality reduction and clustering using an adaptive fuzzy k-menas loss function |
| Geddes et al., 2019 [116] | Cell Clustering | scCCESS | AE | https://github.com/gedcom/scCCESS | Consensus clustering of latent representation clustering from ensemble of random projection or random subset of gene autoencoders |
| Li et al., 2020 [117] | Cell Clustering | DESC | AE | https://github.com/eleozzr/desc | Pretrain stacked autoencoder, then perform simultaneous clustering and dimensionality reduction using deep embedding clustering |
| Lakkis et al., 2021 [118] | Cell Clustering | CarDEC | AE | https://github.com/jlakkis/CarDEC | Separate encoder for high and low expressing genes with separate loss functions to improve single cell representation |
| Chen et al., 2020 [119] | Cell Clustering | scziDesk | AE | https://github.com/xuebaliang/ scziDesk | Weighted soft k-means clustering of latent space during autoencoder training |
| Wang et al., 2021 [120] | Cell Clustering | scGNN | AE \| GAE | https://github.com/juexinwang/scGNN | Combination of several autoencoder structures, including a graph autoencoder to perform entire piepline of single cell analysis after pre-processing |
| Srinivasan et al., 2020 [23] | Cell Clustering | DUSC | DAE | https://github.com/KorkinLab/DUSC | Denoising autoencoder for dimensionality reduction |
| Zeng et al., 2021 (Preprint) [24] | Cell Clustering | GraphSCC | GCN \| DAE | https://github.com/GeniusYx/GraphSCC | Joint residual graph convolutional network and denoising autoencoder with simultaneous clustering for improved latent representation and clustering |
| Amodio et al., 2019 [121] | Cell Clustering | SAUCIE | AE | https://github.com/KrishnaswamyLab/ SAUCIE | Information dimension regularization and cluster distance regularization for improved clustering |
| Li et al., 2021 [122] | Cell Clustering | EMDEC | AE | - | Use optimization procedure for hyperparameters and architecture for DEC with single cell RNA-seq data |
| Kopf et al., 2020 [123] | Cell Clustering | MoE-Sim-VAE | VAE | https://github.com/andkopf/MoESimVAE | Use mixture of gaussians prior for VAE, define separate decoders for each gaussian for reconstruction, and use similarity+DEPICT loss function for clustering |
| Ding et al., 2018 [25] | Cell Clustering | SCVIS | VAE | https://bitbucket.org/jerry00/scvis-dev | Probabilistic generative model with asymmetric t-SNE objective for improved clustering with dimensionality reduction |
| Chen et al., 2020 [127] | Cell Type Annotation | scAnCluster | AE | https://github.com/xuebaliang/scAnCluster | Inclusion of soft k-means clustering with entropy regularization and a self-supervised cell similarity loss for improved clustering |
| Goyal et al., 2020 (Preprint) [128] | Cell Type Annotation | JIND | DFNN | https://github.com/mohit1997/JIND | Use adversarial training to match latent representation coming from source and target domains for downstream cell annotation |
| Hu et al., 2020 [129] | Cell Type Annotation | ItClust | DAE | https://github.com/jianhuupenn/ItClust | Pretrain model on source dataset and then finetune on target dataset |
| Shao et al., 2021 [130] | Cell Type Annotation | scDeepSort | GAE | https://github.com/ZJUFanLab/scDeepSort | Use graph neural network on cell-gene graph to predict pre-defined cell types |
| Li et al., 2020 (Preprint) [131] | Cell Type Annotation | AutoClass | AE | https://github.com/datapplab/AutoClass | Pseudo-labels from k-means clustering or known cell types during training to improve autoencoder-based imputation |
| Xu et al., 2021 [132] | Cell Type Annotation | scANVI | VAE | https://github.com/YosefLab/scvi-tools | Semi-supervised extension of scVI |
| Chen et al., 2020 [133] | Cell Type Annotation | scSemiCluster | AE | https://github.com/xuebaliang/scSemiCluster | Cluster Compactness Loss for labeled data to improve transfer learning |
| Zhou et al., 2021 (Preprint) [134] | Cell Type Annotation | scAdapt | GAN | https://github.com/zhoux85/scAdapt | Use virtual adversarial training loss and semantic alignment loss to improve training in a semi-supervised setting |
| Lotfallahi et al., 2021 [135] | Cell Type Annotation | scArches | VAE | https://github.com/theislab/scarches | Concatenation of new dataset to pretrained autoencoder ("architectural surgery") for improved mapping of query to reference dataset |
| Brbić et al., 2020 [136] | Cell Type Annotation | MARS | AE | https://github.com/snap-stanford/mars | Meta-learning approach to allow for identification of new clusters during transfer learning in new datasets |
| Zhang et al., 2021 [137] | Cell Type Annotation | $MAT^2$ | AE | https://github.com/Zhang-Jinglong/MAT2 | Generate triplets using either known cell labels, or pseudo-labels based on Seurat for contrastive learning using triplet loss and use triplet loss in batch correction |
| Kimmel et al., 2021 [138] | Cell Type Annotation | scNym | DFNN | https://github.com/calico/scnym | Use MixMatch for semi-supervised learning |
| Song et al., 2021 [139] | Cell Type Annotation | scGCN | GCN | https://github.com/QSong-github/scGCN | Development of multiple mutual nearest neighbor graphs based on CCA using reference and query datasets for transfer learning |
| Yuan et al., 2021 [140] | Cell Type Annotation | scMRA | AE \| GCN | https://github.com/ddb-qiwang/scMRA-torch | Development of cell type prototype knowledge graph based on multiple different source domains for improved transfer learning to unlabeled dataset |
| Koh et al., 2021 [141] | Cell Type Annotation | MapCell | DFNN | https://github.com/lianchye/mapcell | Use siamese network with contrastive loss for pairs of cells identified as the same type. Use learned distance metric for label transfer and new cell discovery |
| Wang et al., 2021 [142] | Cell Type Annotation | SigGCN | GAE \| DFNN | https://github. com/NabaviLab/sigGCN | Concatenate latent representation learned from FFNN and GAE to predict cell type |
| Yin et al., 2021 [143] | Cell Type Annotation | scIAE | AE | https://github.com/JGuan-lab/scIAE | Use ensemble of autoencoders with random projections to perform dimensionality reduction. Use the learned representations to train downstream classifiers for new data |
| Duan et al., 2021 [144] | Cell Type Annotation | mtSC | DFNN | https://github.com/bm2-lab/mtSC | Use N-pair loss for deep metric learning across all reference datasets separately for trained model and use a consensus score from each reference dataset for cell annotation of query cell |
| Liu et al., 2021 [145] | Cell Type Annotation | ImmClassifier | DFNN | https://github.com/xliu-uth/ImmClassifier | Use probability of coarse cell predictions into fine-grain predictions using the coarse grain probability distribution as input of a FFNN |
| Dong et al., 2021 [146] | Cell Type Annotation | netAE | VAE | https://github.com/LeoZDong/netAE | Introduction of cell classification on latent representation for labeled cells and modularity loss based on cell-cell similarity matrix of latent representation |
| Cao et al., 2020 [148] | Cell Type Annotation | Cell BLAST | VAE | https://github.com/gao-lab/Cell_BLAST | Use of improved distance-metric for mapping query cell to reference latent-representation and includes poission distribution as method for data augmentation of input scRNA-seq data |
| Wang et al., 2021 [147] | Cell Type Annotation | MultiCapsNet | CapsNet [188] | https://github.com/bojone/Capsule | Use of Capsule Network for single-cell sequencing analysis |
| Du et al. 2020 (Preprint) [151] | Trajectory Analysis | VITAE | VAE | https://github.com/jaydu1/VITAE | Use hierarchical mixture model based on latent representation from VAE to predict cell pseudotime |
| Tangherloni et al., 2021[186] | Complete analysis framework | scAEspy | - | https://gitlab.com/cvejic-group/scaespy | Single-cell analysis package containing several different autoencoder architectures for analysis |
| Fischer et al., 2021 [187] | Complete analysis framework | sfaira | - | https://github.com/theislab/sfaira | Single cell package containing pipeline and pretrained models |
| Abbreviations: AE = autoencoder; CapsNet = capsule neural network; DAE = denoising autoencoder; DFNN = deep feed-forward neural network; GAN = generative adversarial networks; GAE = graph autoencoder; GCN = graph convolutional network; VAE = variational autoencoder. | | | | | |

# Figures

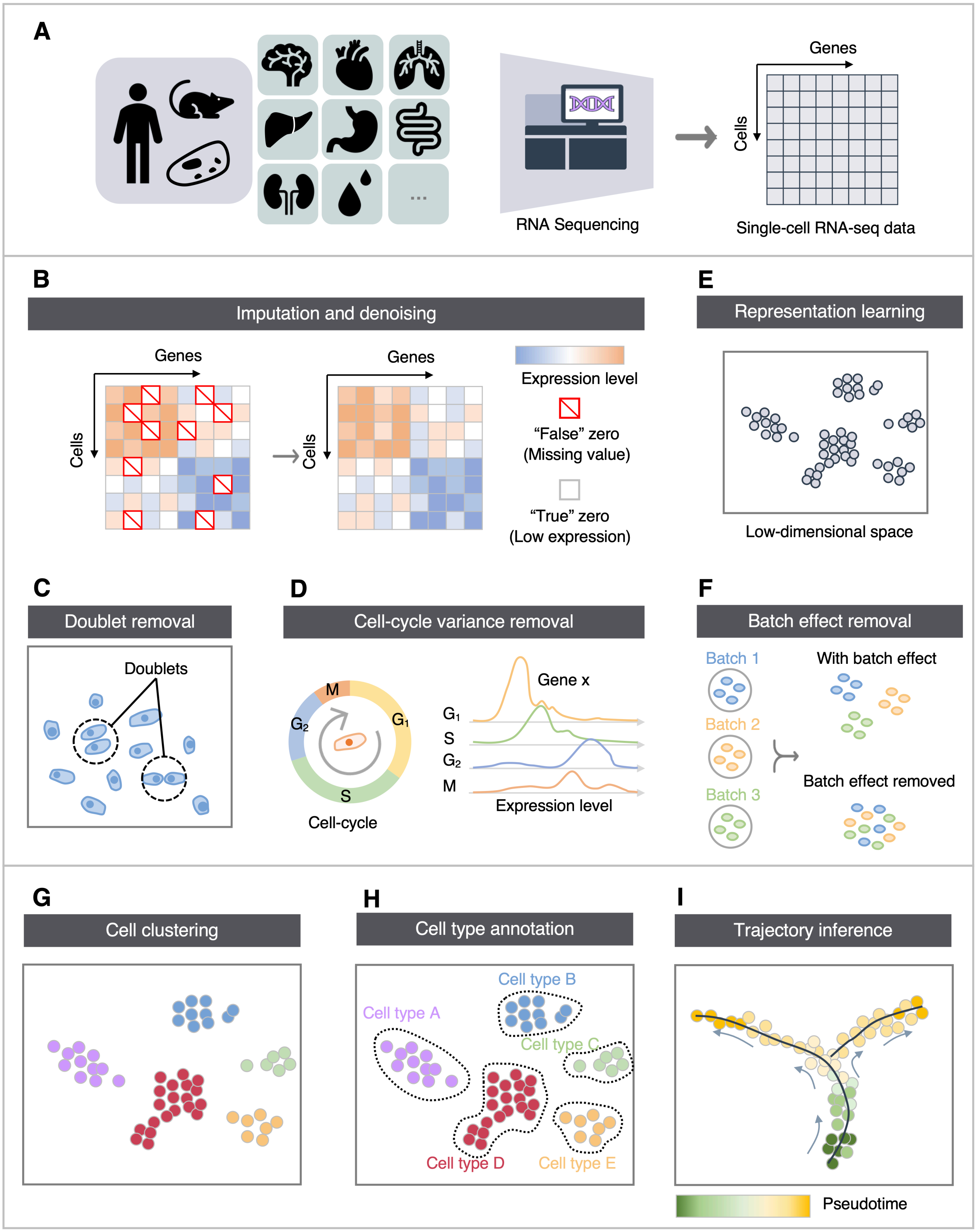

**Figure 1. Schematic of the common pipeline in scRNA-seq analysis. A.** scRNA-seq data collection. **B-F.** scRNA-seq data preprocessing where deep learning has been involved, including imputation and denoising, doublet removal, cell-cycle variance removal, representation learning for dimensionality reduction, and batch effect removal. **G-I.** Downstream analyses of scRNA-seq data, including cell clustering, cell type annotation, and trajectory inference.

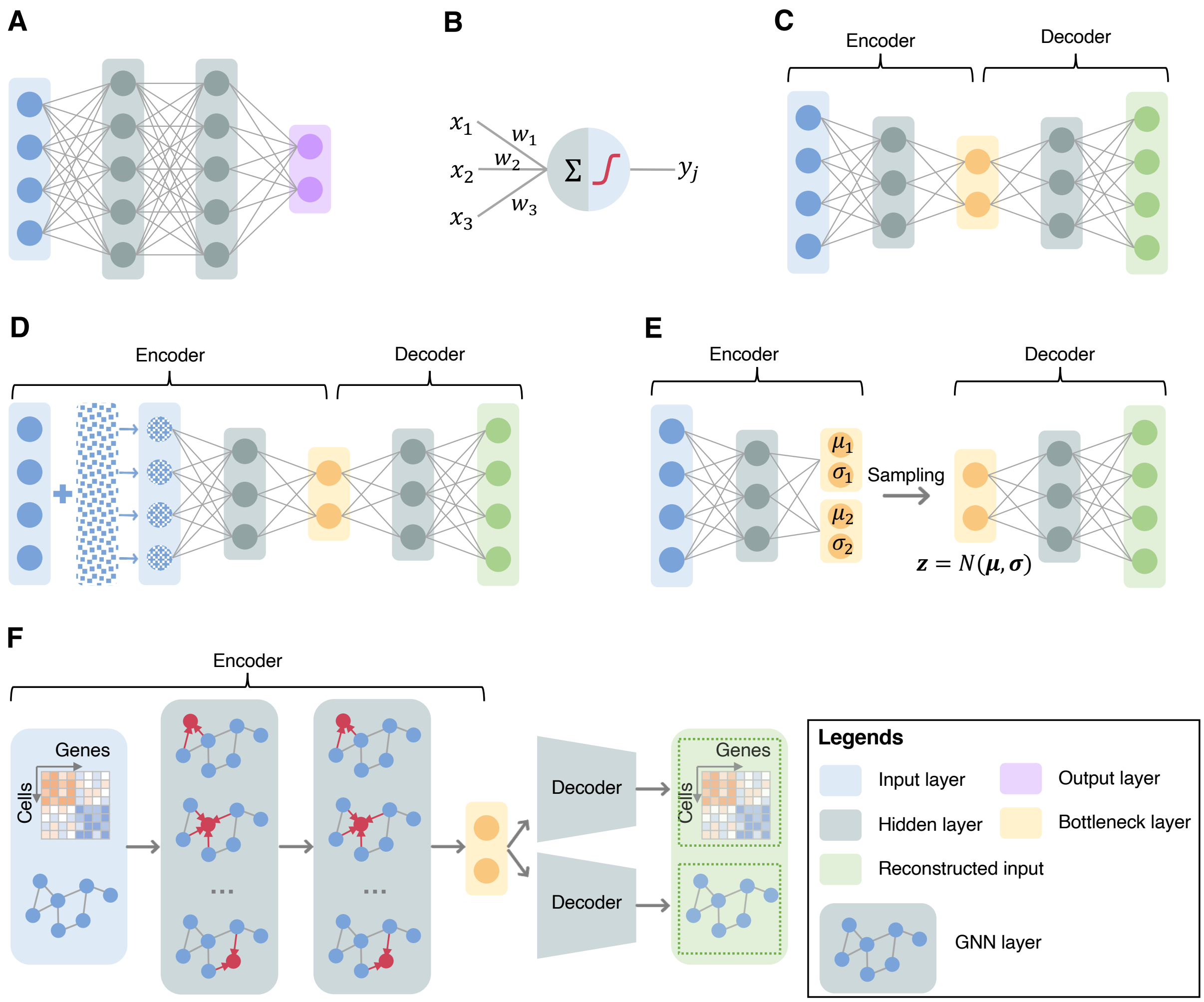

**Figure 2. Illustration of deep learning architectures that have been used in scRNA-seq analysis. A.** Basic design of a feed-forward neural network. **B.** A neural network is composed of "neurons" organized into layers. Each neuron combines a set of weights from the prior layer, passes the weighted summed value through a non-linear activation function, such as sigmoid, rectifier (i.e., rectified linear unit [ReLU]), and hyperbolic tangent, to produce a transformed output. In scRNA-seq data analysis, Autoencoder, a special variant of the feedforward neural network aiming at learning low-dimensional representations of data while preserving data information, has been widely used (**C**). To overcome its pitfalls, variants of Autoencoder have been developed. In order to address the overfitting problems, the Denoising Autoencoder (DAE) forces the input data to be partially corrupted and tries to reconstruct the raw un-corrupted data (**D**); the Variational Autoencoder (VAE) aims at compressing input data into a constrained multivariate latent

distribution space in the encoder, which is regular enough and can be used to generate new content in the decoder (**E**). Benefiting from the advanced deep learning architecture, the Graph Neural Network (GNN), Graph Autoencoder (GAE) has been developed. The encoder of GAE considers both sample features (e.g., cells' gene expression profiles/counts) and samples' neighborhood information (e.g., topological structure of cellular interaction network) to produce low-dimensional representations while preserving topology in data; the decoder unpacks the low-dimensional representations to reconstruct the input network structure and/or sample features (**F**).

Abbreviations: scRNA-seq = single-cell RNA sequencing.